\documentclass[showpacs,preprintnumbers,amsmath,amssymb]{revtex4}
\usepackage{graphicx}
\usepackage{dcolumn}
\usepackage{bm}
\usepackage{amssymb}
\usepackage{array}
\usepackage{hhline}
\usepackage{longtable}
\usepackage[usenames]{color}

\begin{document}

\title{Contact graphs of disk packings as a model of spatial planar networks}

\author{Zhongzhi Zhang$^{1,2}$}
\email{zhangzz@fudan.edu.cn}

\author{Jihong Guan$^{3}$}
\email{jhguan@tongji.edu.cn}

\author{Bailu Ding$^{1,2}$}

\author{Lichao Chen$^{1,2}$}

\author{Shuigeng Zhou$^{1,2}$}
\email{sgzhou@fudan.edu.cn}

\affiliation {$^{1}$School of Computer Science, Fudan University,
Shanghai 200433, China}

\affiliation {$^{2}$Shanghai Key Lab of Intelligent Information
Processing, Fudan University, Shanghai
200433, China} %

\affiliation{$^{3}$Department of Computer Science and Technology,
Tongji University, 4800 Cao'an Road, Shanghai 201804, China}

\begin{abstract}

Spatially constrained planar networks are frequently encountered in
real-life systems. In this paper, based on a space-filling disk
packing we propose a minimal model for spatial maximal planar
networks, which is similar to but different from the model for
Apollonian networks [J. S. Andrade, Jr. et al., Phys. Rev. Lett.
{\bf 94}, 018702 (2005)]. We present an exhaustive analysis of
various properties of our model, and obtain the analytic solutions
for most of the features, including degree distribution, clustering
coefficient, average path length, and degree correlations. The model
recovers some striking generic characteristics observed in most real
networks. To address the robustness of the relevant network
properties, we compare the structural features between the
investigated network and the Apollonian networks. We show that
topological properties of the two networks are encoded in the way of
disk packing. We argue that spatial constrains of nodes are relevant
to the structure of the networks.

\end{abstract}


\date{\today}
\maketitle

\section{Introduction}

The past decade has witnessed a great deal of activity devoted to
complex networks by the scientific community, since many systems in
the real world can be described and characterized by complex
networks~\cite{AlBa02,DoMe02}. Prompted by the computerization of
data acquisition and the increased computing power of computers,
researchers have done a lot of empirical studies on diverse real
networked systems, unveiling the presence of some generic properties
of various natural and manmade networks: power-law degree
distribution $P(k) \sim k^{-\gamma}$ with characteristic exponent
$\gamma$ in the range between 2 and 3~\cite{BaAl99}, small-world
effect including large clustering coefficient and small average path
length (APL)~\cite{WaSt98}, and degree correlations~\cite{Newman02}.
These findings are important for our understanding of the real-life
systems, since they strongly affect almost all aspects of various
dynamical processes taking place on
networks~\cite{Ne03,BoLaMoChHw06,DoGoMe08}.

In order to reproduce or explain the above-mentioned striking common
features of real-life systems, there has been a concerted effort in
the last few years within the physical circle and elsewhere to
develop network models to uncover and understand the complexity of
real systems~\cite{AlBa02,DoMe02}. In addition to the seminal
Watts-Strogatz's (WS) small-world network model~\cite{WaSt98} and
Barab\'asi-Albert's (BA) scale-free network model~\cite{BaAl99}, a
huge variety of models and mechanisms have been proposed to mimic
real-world systems, including initial
attractiveness~\cite{DoMeSa00}, aging and cost~\cite{AmScBaSt00},
fitness model~\cite{BiBa01}, duplication~\cite{ChLuDeGa03}, weight
or traffic driven evolution~\cite{BaBaVe04a,WaWaHuYaOu05},
accelerating growth~\cite{MaGa05,ZhFaZhGu09},
coevolution~\cite{GrBl08}, visibility graph~\cite{LaLuBaLuNu08}, to
name but a few. For reviews, see
Refs.~\cite{AlBa02,DoMe02,Ne03,BoLaMoChHw06}. Although significant
progress has been made in the field of network modeling and has led
to a significant improvement in our understanding of complex
systems, it is still a fundamental task and of current interest to
construct models mimicking real networks and reproducing their
generic properties from different angles.

Most previous network models concentrated on topological aspects and
ignored the geographical effects. In these work, the node position
has no special signification, which is reasonable for some networks
that can be considered as lying in an abstract space, such as
scientific collaboration network~\cite{Ne01a}, metabolic
network~\cite{JeToAlOlBa00} and so forth. However, a plethora of
real networks have well-defined node positions, including the
Internet~\cite{FaFaFa99}, power grid~\cite{AlAlNa04}, airline
networks~\cite{BaBaPaVe04}, to name only a few. This class of
networks is often referred to as geographical or spatial networks,
which is a promising kind of networks, since its geography has a
pivotal influence on the dynamics running on the networks, such as
robustness~\cite{HaMa06}, cascading breakdown~\cite{HuYaYa06},
synchronization~\cite{YiWaWaCh08}, disease
spreading~\cite{WaSaSo02}, among others. In addition to the
geography, some real-world geographical networks are also planar.
Typical examples are street networks~\cite{CaScLaPo06}, electronic
circuit networks~\cite{FeJaSo01}, ant-trail
networks~\cite{BuGaLDKu06}, and neural networks~\cite{Sp03}.
Recently several
models~\cite{WaSaSo02,RoCobeHa02,MaSe02,Ba03,MaMiKo05,GaNe06,KoHaBu08}
have been presented that take the the geographical position into
consideration. However, models for spatial planar networks are much
less, with the exception of space-filling networks---Apollonian
networks~\cite{AnHeAnSi05} that have received a great amount of
attention~\cite{ZhCoFeRo06,ZhRoZh06,ZhChZhFaGuZo08,ViAnHeAn07,OlMoLyAnAl09}.

In the paper, inspired by the disk packing and the construction
method of Apollonian networks~\cite{AnHeAnSi05}, we propose a
spatial planar network, where nodes (located at the center of disks)
correspond to the disks and edges represent contact relation. The
suggested network belongs to a deterministically growing class of
self-similar graphs. On the basis of the recursive construction and
self-similar structure of the network, we determine analytically
many relevant topological properties, such as degree distribution,
clustering coefficient, average path length, as well as degree
correlations. The obtained results show that the network is
simultaneously scale-free, small-world, and disassortative, as
observed in a variety of real networks. Except for the fact that all
nodes of the network have separate spatial positions that encode the
network structure, we also show that the network is not only planar,
but also maximally planar. Note that although Apollonian networks
are also derived from a recursive disk packing, a main subject of
the current paper is to compare the topological features of the
presented network and Apollonian networks, with an aim to address
the effect of the way of disk packings (thus the spatial constrains
of nodes) on the network structure, which is important to the theory
of complex networks.

\begin{figure}
\begin{center}
\includegraphics[width=.7\linewidth,trim=50 20 50 0]{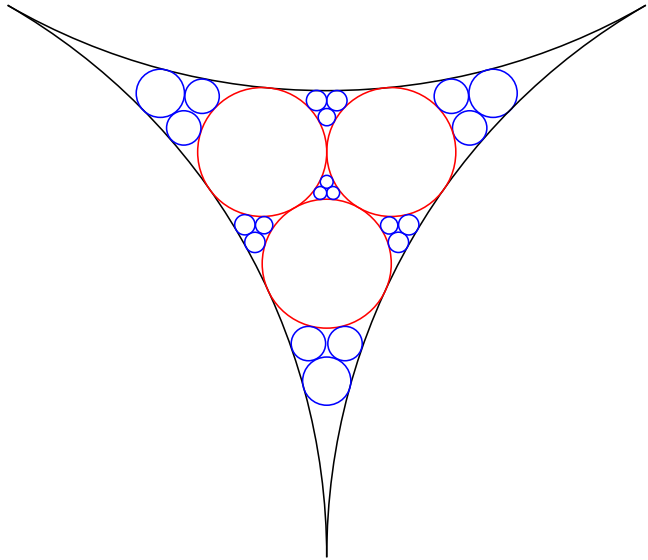}
\end{center}
\caption{\label{packing} The first two generations for the
construction of the disk packing.}
\end{figure}

\section{Construction of the network}

Before defining the network, we first introduce a disk
packing~\cite{HeMaBe90,MaHe91}, which is a variation of the
celebrated Apollonian packing~\cite{Bo73}. The introduced disk
packing, shown in figure~\ref{packing}, is constructed as follows.
We start with three mutually touching disks, the interstice of which
is a curvilinear triangle, and we denote this initial configuration
by generation $t=0$. Then in the first generation $t=1$, three
smaller mutually touching disks are added to fill the interstice of
the initial configuration, each of which touches two adjacent sides
of the initial curvilinear triangle, giving rise to seven new
smaller curvilinear triangles. For subsequent generations we
indefinitely repeat the packing process for all the new curvilinear
triangles. In the limit of infinite $t$ generations, we obtain a
disk packing.

\begin{figure}
\begin{center}
\includegraphics[width=.5\linewidth,trim=50 40 50 0]{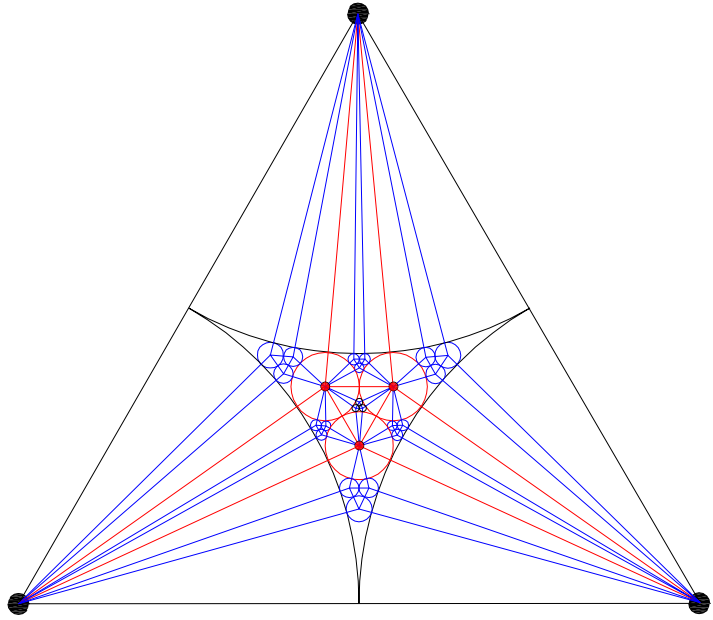}
\end{center}
\caption{\label{net} Illustration of the network corresponding to
the disk packing shown in figure~\ref{packing}.}
\end{figure}

The above-mentioned disk packing can be used as a basis for a
network, which is the research object of current paper. The
translation from the disk packing to network generation is quite
straightforward. Let the nodes (vertices) of the network correspond
to the disks and make two nodes connected if the corresponding disks
are in contact. Alternatively, one can also connect the centers of
the touching disks by lines to obtain the network. Figure~\ref{net}
shows the network corresponding to the disk packing in
figure~\ref{packing}.

In the construction process of the disk packing, for each interstice
at arbitrary generation, once we add three disks to fill it, seven
new interstices are created that will be filled in the next
iteration. When building the network, it is equivalent to say that
for each group of three new nodes added, seven new triangles are
generated in the network, into each of which a group of three nodes
will be inserted in the next iteration. According to this, we can
introduce an iterative algorithm to create the network, denoted by
$F_t$ after $t$ generation evolutions.

\begin{figure}
\begin{center}
\includegraphics[width=0.3\linewidth,trim=100 150 100
80]{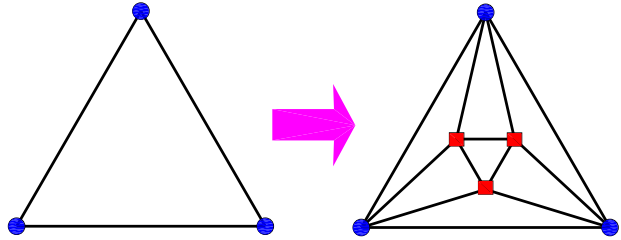}
\end{center}
\caption[kurzform]{\label{iterative} Iterative construction method
for the network. }
\end{figure}

The iterative algorithm for the network is as follows: For $t=0$,
$F_0$ consists of three nodes forming a triangle. Then, we add three
nodes into the original triangle. These three new nodes are linked
to each other shaping a new triangle, and both ends of each edge of
the new triangle are connected to a node of the original triangle.
Thus we get $F_1$, see figure~\ref{iterative}. For $t\geq 1$, $F_t$
is obtained from $F_{t-1}$. For each of the existing triangles of
$F_{t-1}$ that has never generated nodes before, we call it an
active triangle. We replace each of the existing active triangles of
$F_{t-1}$ by the connected cluster on the right hand side of
figure~\ref{iterative} to obtain $F_t$. The growing process is
repeated until the network reaches a desired order (node number of
network). Figure~\ref{net} shows the network growing process for the
first two steps.

Next we compute the order and size (number of all edges) of the
network $F_t$. Let $L_v(t)$, $L_e(t)$ and $L_\Delta(t)$ be the
number of nodes, edges and active triangles created at step $t$,
respectively. By construction (see also figure~\ref{iterative}),
each active triangle in $F_{t-1}$ will be replaced by seven active
triangles in $F_t$. Thus, it is not difficult to find the following
relation: $L_\Delta(t)=7\,L_\Delta(t-1)$. Since $L_\Delta(0)=1$, we
have $L_\Delta(t)=7^{t}$.

Note that each active triangle in $F_{t-1}$ will lead to an addition
of three new nodes and nine new edges at step $t$, then one can
easily obtain the following relations:
$L_v(t)=3\,L_\Delta(t-1)=3\times7^{t-1}$, and
$L_e(t)=9\,L_\Delta(t-1)=9\times7^{t-1}$ for arbitrary $t>0$. From
these results, we can compute the order and size of the network. The
total number of vertices $N_t$ and edges $E_t$ present at step $t$
is
\begin{equation}\label{Nt}
N_t=\sum_{t_i=0}^{t}L_v(t_i)=\frac{7^{t}+5}{2}
\end{equation}
and
\begin{equation}\label{Et}
E_t =\sum_{t_i=0}^{t}L_e(t_i)=\frac{3\times7^{t}+3}{2},
\end{equation}
respectively. So for large $t$, the average degree $\overline{k}_t=
\frac{2E_t}{N_t}$ is approximately $6$, which shows the network is
sparse as most real systems.

From equations.~(\ref{Nt}) and~(\ref{Et}), we have $E_t=3N_{t}-6$.
In addition, by the very construction of the network, it is obvious
that arbitrary two edges in the network never cross each other. Thus
our network is a maximal planar network (or graph)~\cite{We01}.

\section{Structural properties of the network}

In this section, we study the statistical properties of the network,
in terms of degree distribution, clustering coefficient, average
path length, and degree correlations.

\subsection{Degree distribution}

When a new node $i$ is added to the graph at step $t_i$ ($t_i\geq
1$), it has a degree of $4$. Let $L_\Delta(i,t)$ be the number of
active triangles at step $t$ that will create new nodes connected to
node $i$ at step $t+1$. Then at step $t_i$, $L_\Delta(i, t_i)=4$.
From the iterative generation process of the network, one can see
that at any step each two new neighbors of $i$ generate three new
active triangles involving $i$, and one of its existing active
triangle is deactivated simultaneously. We define $k_i(t)$ as the
degree of node $i$ at time $t$, then the relation between $k_i(t)$
and $L_\Delta(i,t)$ satisfies:
\begin{equation}\label{deltak}
L_\Delta(i,t)=k_i(t).
\end{equation}
Now we compute $L_\Delta(i,t)$. By construction,
$L_\Delta(i,t)=3\,L_\Delta(i,t-1)$. Considering the initial
condition $L_\Delta(i, t_i)=4$, we can derive
$L_\Delta(i,t)=4\times3^{t-t_{i}}$. Then at time $t$, the degree of
vertex $i$ becomes
\begin{equation}\label{ki}
k_i(t)=4\times3^{t-t_{i}}.
\end{equation}

It should be mentioned that the initial three vertices created at
step 0 have a little different evolution process from other ones. We
can easily obtain: $L_\Delta(0,t)=3^{t}$ and $k_i(t)=3^{t}+1$. Thus,
at step $t$, the degree of the initial three nodes is less than that
of those three nodes born at step 1 but larger than that for those
nodes emerging at other steps.

Equation~(\ref{ki}) shows that the degree spectrum of the network is
discrete. It follows that the cumulative degree
distribution~\cite{Ne03} is given by
\begin{equation}\label{pcumk}
P_{\rm cum}(k)=\sum_{\tau \leq t_i}\frac{L_v(\tau)}{N_t}
={7^{t_i}+5\over 7^{t}+5},
\end{equation}
which is valid for all $t_i\geq2$.

Substituting for $t_i$ ($t_i\geq2$) in this expression using
$t_i=t-\frac{\ln(\frac{k}{4})}{\ln 3}$ gives
\begin{equation}
P_{\rm
cum}(k)=\frac{7^{t}\times(\frac{k}{4})^{-(\ln7/\ln3)}+5}{7^{t}+5}.
\end{equation}
When $t$ is large enough, one can obtain
\begin{equation}\label{gammak}
P_{\rm cum}(k)=\left(\frac{k}{4}\right)^{-\ln7/\ln3}.
\end{equation}
So the degree distribution follows a power law form with the
exponent $\gamma=1+\frac{\ln7}{\ln3}$.

\subsection{Clustering coefficient}

The clustering coefficient~\cite{WaSt98} $ C_i $ of node $i$ is
defined as the ratio between the number of edges $e_i$ that actually
exist among the $k_i $ neighbors of node $i$ and its maximum
possible value, $ k_i( k_i -1)/2 $, i.e., $ C_i =2e_i/[k_i( k_i
-1)]$. The clustering coefficient of the whole network is the
average of $C_i^{'}s $ over all nodes in the network.

For our network, the analytical expression of clustering coefficient
$C(k)$ for a single node with degree $k$ can be derived exactly.
When a node enters the system, both $k_i$ and $e_i$ are 4. In the
following iterations, each of its active triangles increases both
$k_{i}$ and $e_{i}$ by 2 and 3, respectively. Thus, $e_{i}$ is equal
to $4+\frac{3}{2}\left(k_{i}-4\right)$ for all nodes at all steps.
So one can see that there exists a one-to-one correspondence between
the degree of a node and its clustering. For a node of degree $k$,
we have
\begin{equation}\label{Ck}
C(k)= \frac{2\,e}{k(k-1)}=
\frac{2\left[4+\frac{3}{2}(k-4)\right]}{k(k-1)}=\frac{4}{k}-\frac{1}{k-1}.
\end{equation}
In the limit of large $k$, $C(k)$ is inversely proportional  to
degree $k$. The same scaling of $C(k)\sim k^{-1}$ has also been
observed in several real-life networks~\cite{RaBa03}.

\begin{figure}
\begin{center}
\includegraphics[width=.50\linewidth,trim=50 40 50 0]{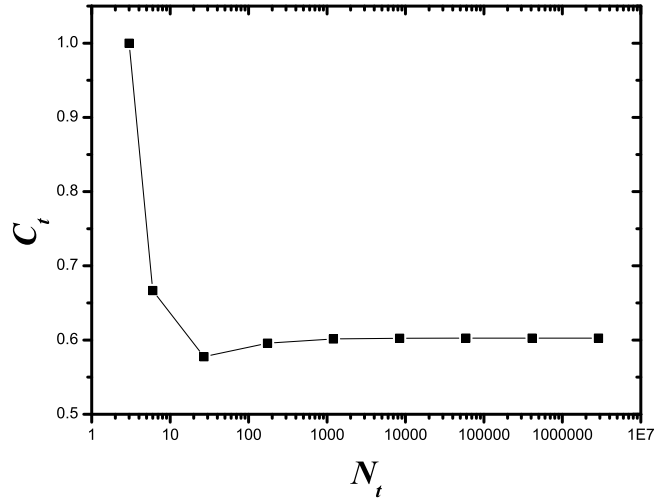}
\end{center}
\caption[kurzform]{\label{clustering} Semilogarithmic plot of
average clustering coefficient $C_t$ versus network order $N_{t}$.}
\end{figure}

Using equation (\ref{Ck}), we can obtain the clustering $C_t$ of the
network at step $t$:
\begin{equation}\label{ACC}
C_t=
    \sum_{r=0}^{t}\left[\frac{
    L_v(r)}{N_{t}}\left(\frac{4}{D_r}-\frac{1}{D_r-1}\right)\right],
\end{equation}
where the sum runs over all the nodes and $D_r$ is the degree of the
nodes created at step $r$, which is given by equation~(\ref{ki}). In
the infinite network order limit ($N_{t}\rightarrow \infty$),
equation~(\ref{ACC}) converges to a nonzero value $C=0.603$, as
shown in figure~\ref{clustering}. Therefore, the average clustering
coefficient of the network is very high.

\subsection{Average path length}

We represent all the shortest path lengths of network $F_{t}$ as a
matrix in which the entry $d_{ij}$ is the distance between node $i$
and $j$ that is the length of a shortest path joining $i$ and $j$. A
measure of the typical separation between two nodes in $F_{t}$ is
given by the average path length (also called average distance)
$d_{t}$ defined as the mean of distances over all pairs of nodes:
\begin{equation}\label{apl01}
d_{t} = \frac{D_t}{N_t(N_t-1)/2}\,,
\end{equation}
where
\begin{equation}\label{total01}
D_t = \sum_{i \in F_{t},\, j \in F_{t},\, i \neq j} d_{ij}
\end{equation}
denotes the sum of the distances between two nodes over all couples.
Note that in equation~(\ref{total01}), for a pair of nodes $i$ and
$j$ ($i \neq j$), we only count $d_{ij}$ or $d_{ji}$, not both.

\begin{figure}
\begin{center}
\includegraphics[width=0.5\linewidth,trim=100 10 100 10]{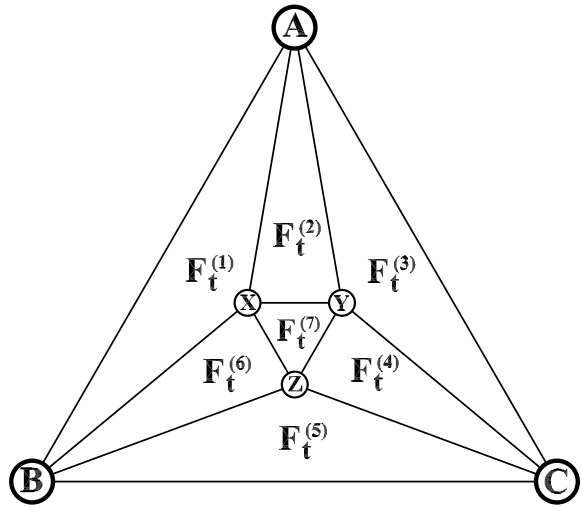} \caption{Schematic illustration of second construction
means of the network. $F_{t+1}$ may be obtained by joining seven
copies of $F_t$ denoted as
    $F_t^{(\eta)}$ $(\eta=1,2,\cdots,7)$, which are
    connected to one another at the six edge nodes, i.e., $A$, $B$, $C$, $X$, $Y$, and $Z$.} \label{copy}
\end{center}
\end{figure}

\subsubsection{Recursive equation for total distances}

We continue by exhibiting the procedure of determining the total
distance and present the recurrence formula, which allows us to
obtain $D_{t+1}$ of the $t+1$ generation from $D_{t}$ of the $t$
generation. The network $F_{t}$ under consideration has a
self-similar structure that allows one to calculate $D_{t}$
analytically~\cite{HiBe06,ZhZhCh07,ZhZhWaSh07}. As shown in
figure~\ref{copy}, network $F_{t+1}$ may be constructed by joining
at six edge nodes (i.e., $A$, $B$, $C$, $X$, $Y$, and $Z$) seven
copies of $F_{t}$ that are labeled as $F_{t}^{(1)}$, $F_{t}^{(2)}$,
$\cdots$, $F_{t}^{(7)}$.

According to the second construction method, the total distance
$D_{t+1}$ satisfies the recursion relation
\begin{equation}\label{total02}
  D_{t+1} = 7\, D_t + \Delta_t-9,
\end{equation}
where $\Delta_t$ is the sum over all shortest path length whose
endpoints are not in the same $F_t^{(\eta)}$ branch. The last term
-9 on the right-hand side of equation~(\ref{total02}) compensates
for the overcounting of certain paths: the shortest path $d_{AX}$
between $A$ and $X$, with length 1, is included in both
$F_{t}^{(1)}$ and $F_{t}^{(2)}$; similarly, the shortest paths
$d_{AY}$, $d_{BX}$, $d_{BZ}$, $d_{CY}$, $d_{CZ}$, $d_{XY}$,
$d_{XZ}$, and $d_{YZ}$ are all computed twice. To determine $D_t$,
all that is left is to calculate $\Delta_t$.

\subsubsection{Definition of crossing distance}

In order to compute $\Delta_t$, we classify the nodes in $F_{t+1}$
into two categories: the six edge nodes (such as $A$, $B$, $C$, $X$,
$Y$, and $Z$ in figure~\ref{copy}) are called connecting nodes,
while the other nodes are named non-connecting nodes. Thus
$\Delta_t$, named the crossing distance, can be obtained by summing
the following path length that are not included in the distance of
node pairs in $F_{t}^{(\eta)}$: length of the shortest paths between
non-connecting nodes, length of the shortest paths between
connecting and non-connecting nodes, and length of the shortest
paths between connecting nodes (i.e., $d_{AZ}$, $d_{BY}$, and
$d_{CZ}$).

Denote $\Delta_t^{\alpha,\beta}$ as the sum of all shortest paths
between non-connecting nodes, whose end-points are in
$F_t^{(\alpha)}$ and $F_t^{(\beta)}$, respectively. That is to say,
$\Delta_t^{\alpha,\beta}$ rules out the paths with end-point at the
connecting nodes belonging to $F_t^{(\alpha)}$ or $F_t^{(\beta)}$.
For example, each path contributed to $\Delta_t^{1,2}$ does not end
at node $A$, $B$, $X$ or $Y$. On the other hand, let
$\Omega_t^{\eta}$ be the set of non-connecting nodes in $F_{t+1}$,
which belong to $F_{t}^{(\eta)}$. Then the total sum $\Delta_t$ is
given by
\begin{eqnarray}\label{cross01}
\Delta_t
=\sum_{\beta=\alpha+1}^{7}\sum_{\alpha=1}^{6}\Delta_t^{\alpha,\beta}&+&
\sum_{\stackrel{j \in \Omega_t^{\eta}}{\eta \in
      \{4,5,6,7\}}}d_{Aj}+
\sum_{\stackrel{j \in \Omega_t^{\eta}}{\eta \in
      \{2,3,4,7\}}}d_{Bj}+
\sum_{\stackrel{j \in \Omega_t^{\eta}}{\eta \in
      \{1,2,6,7\}}}d_{Cj} +
\sum_{\stackrel{j \in \Omega_t^{\eta}}{\eta \in
      \{3,4,5\}}}d_{Xj}\nonumber \\
      &+&
\sum_{\stackrel{j \in \Omega_t^{\eta}}{\eta \in
      \{1,5,6\}}}d_{Yj}+
\sum_{\stackrel{j \in \Omega_t^{\eta}}{\eta \in
      \{1,2,3\}}}d_{Zj}+d_{AZ}+d_{BY}+d_{CX}
\end{eqnarray}

By symmetry, equation~(\ref{cross01}) can be simplified as
\begin{equation}\label{cross02}
\Delta_t
=9\,\Delta_t^{1,2}+9\,\Delta_t^{1,3}+3\,\Delta_t^{1,4}+21\,\sum_{j
\in \Omega_t^{4}}d_{Aj}+6.
\end{equation}
Having $\Delta_t$ in terms of the quantities of $\Delta_t^{1,2}$,
$\Delta_t^{1,3}$, $\Delta_t^{1,4}$, and $\sum_{j \in
\Omega_t^{4}}d_{Aj}$, the next step is to explicitly determine these
quantities.

\begin{figure}
\begin{center}
\includegraphics[width=0.55\linewidth,trim=100 10 100
0]{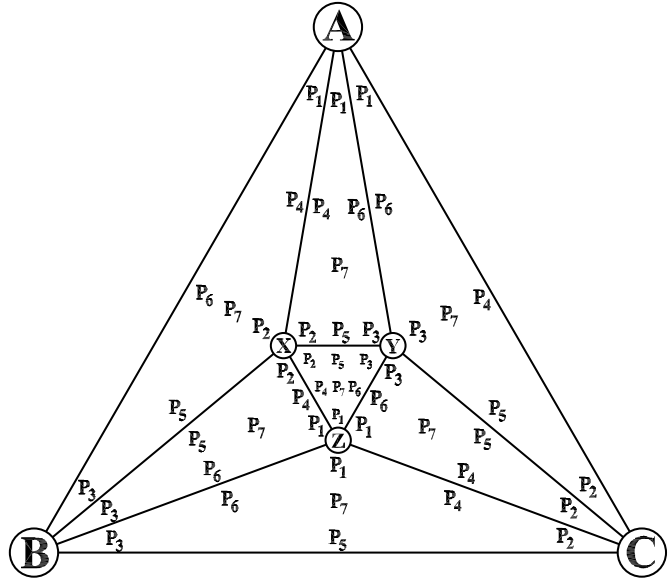} \caption{Illustration of the classification of
interior nodes in $F_t^{(\eta)}$ $(\eta=1,2,\cdots,7)$, from which
we can derive recursively the classification of interior nodes in
network $F_{t+1}$.}\label{class}
\end{center}
\end{figure}

\subsubsection{Classification of interior nodes}

To calculate the crossing distance $\Delta_t^{1,2}$,
$\Delta_t^{1,3}$, $\Delta_t^{1,4}$, and $\sum_{j \in
\Omega_t^{4}}d_{Aj}$, we classify interior nodes in network
$F_{t+1}$ into seven different parts according to their shortest
path lengths to each of the three peripheral nodes (i.e. $A$, $B$,
$C$). Notice that nodes $A$, $B$, $C$ themselves are not partitioned
into any of the seven parts represented as $P_{1}$, $P_{2}$,
$P_{3}$, $P_{4}$, $P_{5}$, $P_{6}$, and $P_{7}$, respectively. The
classification of nodes is shown in figure~\ref{class}. For any
interior node $v$, we denote the shortest path lengths from $v$ to
$A$, $B$, $C$ as $a$, $b$, and $c$, respectively. By construction,
$a$, $b$, $c$ can differ by at most $1$ since vertices $A$, $B$, $C$
are adjacent. Then the classification function $class(v)$ of node
$v$ is defined to be

\begin{equation}\label{classification}
class(v)=\left\{
\begin{array}{lc}
{\displaystyle{P_{1}}}
& \quad \hbox{for}\ a<b=c,\\
{\displaystyle{P_{2}}}
& \quad \hbox{for}\ b<a=c,\\
{\displaystyle{P_{3}}}
& \quad \hbox{for}\ c<a=b,\\
{\displaystyle{P_{4}}}
& \quad \hbox{for}\ a=c<b,\\
{\displaystyle{P_{5}}}
& \quad \hbox{for}\ a=b<c,\\
{\displaystyle{P_{6}}}
& \quad \hbox{for}\ b=c<a,\\
{\displaystyle{P_{7}}}
& \quad \hbox{for}\ a=b=c.\\
\end{array} \right.
\end{equation}

It should be mentioned that the definition of node classification is
recursive. For instance, class $P_{1}$ and $P_{4}$ in $F_t^{(1)}$
belong to class $P_{1}$ in $F_{t+1}$, class $P_{3}$ and $P_{5}$ in
$F_t^{(1)}$ belong to class $P_{2}$ in $F_{t+1}$, class $P_{2}$,
$P_{6}$, and $P_{7}$ in $F_t^{(1)}$ belong to class $P_{5}$ in
$F_{t+1}$. Since the three nodes $A$, $B$, and $C$ are symmetrical,
in the network we have the following equivalent relations from the
viewpoint of class cardinality: classes $P_{1}$, $P_{2}$, and
$P_{3}$ are equivalent to one another, and it is the same with
classes $P_{4}$, $P_{5}$, and $P_{6}$. We denote the number of nodes
in network $F_{t}$ that belong to class $P_{1}$ as $N_{t,P_{1}}$,
the number of nodes in class $P_{2}$ as $N_{t,P_{2}}$, and so on. By
symmetry, we have $N_{t,P_{1}}=N_{t,P_{2}}=N_{t,P_{3}}$ and
$N_{t,P_{4}}=N_{t,P_{5}}=N_{t,P_{6}}$. Therefore in the following
computation we will only consider $N_{t,P_{1}}$, $N_{t,P_{4}}$, and
$N_{t,P_{7}}$. It is easy to conclude that
\begin{align}
N_{t} &= N_{t,P_{1}}+N_{t,P_{2}}+N_{t,P_{3}}+N_{t,P_{4}}+N_{t,P_{5}}+N_{t,P_{6}}+N_{t,P_{7}}+3 \nonumber\\
 &=3\,N_{t,P_{1}} + 3\,N_{t,P_{4}} + N_{t,P_{7}}+3.
\end{align}
Considering the self-similar structure of the network, we can easily
know that at time $t+1$, the quantities $N_{t+1,P_{1}}$,
$N_{t+1,P_{4}}$, and $N_{t+1,P_{7}}$ evolve according to the
following recursive equations
\begin{eqnarray}\label{Np01}
\left\{
\begin{array}{ccc}
N_{t+1,P_{1}} &=& 3\,N_{t,P_{1}} + 4\,N_{t,P_{4}}+N_{t,P_{7}}\,, \\
N_{t+1,P_{4}} &=& 4\,N_{t,P_{1}} + N_{t,P_{4}}+N_{t,P_{7}}+1\,, \\
N_{t+1,P_{7}} &=& 6\,N_{t,P_{4}}+N_{t,P_{7}}\,, \\
 \end{array}
 \right.
\end{eqnarray}
where we have used the equivalent relations
$N_{t,P_{1}}=N_{t,P_{2}}=N_{t,P_{3}}$ and
$N_{t,P_{4}}=N_{t,P_{5}}=N_{t,P_{6}}$.  With the initial condition
$N_{2,P_{1}}=4$, $N_{2,P_{4}}=2$, and $N_{2,P_{7}}=6$, we can solve
the recursive equation~(\ref{Np01}) to obtain

\begin{eqnarray}\label{Np02}
\left\{
\begin{array}{ccc}
N_{t,P_{1}} &=& \frac{1}{124} \left [-62+10\cdot7^t+26
\left(-1-\sqrt{2}\right)^t+26 \left(-1+\sqrt{2}\right)^t+11
\sqrt{2} \left(-1+\sqrt{2}\right)^t-11 \sqrt{2} \left(-1-\sqrt{2}\right)^t\right ]\,, \\
N_{t,P_{4}} &=& \frac{1}{124} \left[8\cdot7^t-4
\left(-1-\sqrt{2}\right)^t-4 \left(-1+\sqrt{2}\right)^t+15
\sqrt{2} \left(-1+\sqrt{2}\right)^t-15 \sqrt{2} \left(-1-\sqrt{2}\right)^t\right], \\
N_{t,P_{7}} &=& \frac{1}{62} \left [62+4\cdot7^t-33
\left(-1-\sqrt{2}\right)^t-33 \left(-1+\sqrt{2}\right)^t+39 \sqrt{2}
\left(-1-\sqrt{2}\right)^t-39
\sqrt{2} \left(-1+\sqrt{2}\right)^t\right]. \\
\end{array}
\right.
\end{eqnarray}

For a node $v$ in network $F_{t+1}$, we are also interested in the
smallest value of the shortest path length from $v$ to any of the
three peripheral nodes $A$, $B$, and $C$. We denote the shortest
distance as $f_v$, which can be defined to be
\begin{equation}\label{fv01}
f_v = min(a,b,c).
\end{equation}

Let $d_{t,P_{1}}$ denote the sum of $f_v$ of all nodes belonging to
class $P_{1}$ in network $F_{t}$. Analogously, we can also define
the quantities $d_{t,P_{2}}$, $d_{t,P_{3}}$, $\cdots$,
$d_{t,P_{7}}$. Again by symmetry, we have
$d_{t,P_{1}}=d_{t,P_{2}}=d_{t,P_{3}}$,
$d_{t,P_{4}}=d_{t,P_{5}}=d_{t,P_{6}}$, and $d_{t,P_{1}}$,
$d_{t,P_{4}}$, $d_{t,P_{7}}$ can be written recursively as follows:

\begin{eqnarray}\label{dp01}
\left\{
\begin{array}{ccc}
d_{t+1,P_{1}} &=& 3\,d_{t,P_{1}} +4\, d_{t,P_{4}}+ d_{t,P_{7}}\,, \\
d_{t+1,P_{4}} &=& 4\,d_{t,P_{1}}+d_{t,P_{4}} + d_{t,P_{7}}+4\,N_{t,P_{1}}+1\,, \\
d_{t+1,P_{7}} &=& 6\,(d_{t,P_{4}} + N_{t,P_{4}})+(d_{t,P_{7}} + N_{t,P_{7}})\,. \\
 \end{array}
 \right.
\end{eqnarray}

Substituting equation~(\ref{Np02}) into equation~(\ref{dp01}), and
considering the initial condition $d_{2,P_{1}}=4$, $d_{2,P_{4}}=2$,
and $d_{2,P_{7}}=12$, equation~(\ref{dp01}) is solved inductively
\begin{eqnarray}\label{dp02}
\left\{
\begin{array}{ccc}
d_{t,P_{1}}&=& \frac{1}{5004888}\Big [7 \left(-3
\left(-1+\sqrt{2}\right)^t \left(20082+18235
\sqrt{2}\right)+\left(-1-\sqrt{2}\right)^t \left(-60246+54705
\sqrt{2}\right)+4 \left(29791+332 7^t\right)\right)\\ &\quad& +93
\left(1760\times7^t-7 \left(-1+\sqrt{2}\right)^t \left(1256+1149
\sqrt{2}\right)+\left(-1-\sqrt{2}\right)^t
\left(-8792+8043 \sqrt{2}\right)\right) t\Big],\\
d_{t,P_{4}}&=& \frac{1}{5004888}\Big [14 \left(-59582+13700\times
7^t+\left(22941-56145 \sqrt{2}\right) \left(-1-\sqrt{2}\right)^t+3
\left(-1+\sqrt{2}\right)^t \left(7647+18715 \sqrt{2}\right)\right)\\
&\quad&+93 \left(1408\times7^t-7 \left(-1+\sqrt{2}\right)^t
\left(1042+107 \sqrt{2}\right)+\left(-1-\sqrt{2}\right)^t
\left(-7294+749 \sqrt{2}\right)\right) t\Big],\\
d_{t,P_{7}}&=& \frac{1}{5004888}\Big [7 \left(9
\left(-1-\sqrt{2}\right)^t \left(-16540+33393 \sqrt{2}\right)-9
\left(-1+\sqrt{2}\right)^t \left(16540+33393 \sqrt{2}\right)+8
\left(29791+7424\times7^t\right)\right)\\ &\quad&+186
\left(704\times7^t-21 \left(-1-\sqrt{2}\right)^t \left(-1149+628
\sqrt{2}\right)+21 \left(-1+\sqrt{2}\right)^t \left(1149+628
\sqrt{2}\right)\right) t\Big].
 \end{array}
 \right.
\end{eqnarray}

\subsubsection{Calculation of crossing distances}

Having obtained the quantities $N_{t,P_{i}}$ and $d_{t,P_{i}}$
($i=1,2,\cdots, 7$), we now begin to determine the crossing distance
$\Delta_t^{1,2}$, $\Delta_t^{1,3}$, $\Delta_t^{1,4}$, and $\sum_{j
\in \Omega_t^{4}}d_{Aj}$ expressed as a function of $N_{t,P_{i}}$
and $d_{t,P_{i}}$. Here we only give the computation details of
$\Delta_t^{1,2}$, while the computing processes of $\Delta_t^{1,3}$,
$\Delta_t^{1,4}$, and $\sum_{j \in \Omega_t^{4}}d_{Aj}$ are similar.
For convenience of computation, we use $\Gamma_t^{\eta,i}$ to denote
the set of interior nodes belonging to class $P_i$ in
$F_{t}^{(\eta)}$. Then $\Delta_t^{1,2}$ can be written as
\begin{equation}\label{cross04}
  \Delta_t^{1,2} = \sum_{\stackrel{u \in \Gamma_t^{1,i},\,i \in \{1,2,3,4,5,6,7\}}{v\in
      F_t^{(2)},\, v \ne A, X, Y }} d_{uv}.
\end{equation}
The seven terms on the right-hand side of equation~(\ref{cross04})
are represented consecutively as $\delta_t^i$ ($i=1,2,\cdots, 7$).
Next we will calculate the quantities $\delta_t^i$. By symmetry,
$\delta_t^1=\delta_t^2$, $\delta_t^5=\delta_t^6$. Therefore, we need
only to compute $\delta_t^1$, $\delta_t^3$, $\delta_t^4$,
$\delta_t^5$ and $\delta_t^7$. Firstly, we evaluate $\delta_t^1$. By
definition,
\begin{eqnarray}\label{cross05}
  \delta_t^1 &=&\sum_{\stackrel{u \in \Gamma_t^{1,1},\,v\in
      F_t^{(2)}}{ v \ne A, X, Y }} d_{uv}\nonumber \\
      &=&\sum_{\stackrel{u \in \Gamma_t^{1,1},v\in
      \Gamma_t^{2,i}}{i \in \{1,4,6,7\} }} (d_{uA}+d_{Av})+\sum_{\stackrel{u \in \Gamma_t^{1,1}}{v\in
      \Gamma_t^{2,3} }}(d_{uA}+d_{AY}+d_{Yv})+\sum_{\stackrel{u \in \Gamma_t^{1,1}}{ v\in
      \Gamma_t^{2,2}\bigcup
      \Gamma_t^{2,5}}}(d_{uA}+d_{AX}+d_{Xv})\nonumber \\
      &=&N_{t,P_{1}}(3d_{t,P_{1}}+3d_{t,P_{4}}+d_{t,P_{7}}+2N_{t,P_{1}}+N_{t,P_{4}})+d_{t,P_{1}}(3N_{t,P_{1}}+3N_{t,P_{4}}+N_{t,P_{7}}).
\end{eqnarray}
Proceeding similarly, we obtain
\begin{eqnarray}\label{cross06}
  \delta_t^3
  =N_{t,P_{1}}(3d_{t,P_{1}}+3d_{t,P_{4}}+d_{t,P_{7}}+4N_{t,P_{1}}+3N_{t,P_{4}}+N_{t,P_{7}})+d_{t,P_{1}}(3N_{t,P_{1}}+3N_{t,P_{4}}+N_{t,P_{7}}),
\end{eqnarray}
\begin{eqnarray}\label{cross07}
  \delta_t^4
  =N_{t,P_{2}}(3d_{t,P_{1}}+3d_{t,P_{4}}+d_{t,P_{7}}+N_{t,P_{1}})+d_{t,P_{2}}(3N_{t,P_{1}}+3N_{t,P_{4}}+N_{t,P_{7}}),
\end{eqnarray}
\begin{eqnarray}\label{cross08}
  \delta_t^5
  =N_{t,P_{2}}(3d_{t,P_{1}}+3d_{t,P_{4}}+d_{t,P_{7}}+2N_{t,P_{1}}+N_{t,P_{4}})+d_{t,P_{2}}(3N_{t,P_{1}}+3N_{t,P_{4}}+N_{t,P_{7}}),
\end{eqnarray}
and
\begin{eqnarray}\label{cross09}
  \delta_t^7
  =N_{t,P_{3}}(3d_{t,P_{1}}+3d_{t,P_{4}}+d_{t,P_{7}}+N_{t,P_{1}})+d_{t,P_{7}}(3N_{t,P_{1}}+3N_{t,P_{4}}+N_{t,P_{7}}).
\end{eqnarray}
With the obtained results for $\delta_t^i$, we have
\begin{eqnarray}\label{cross10}
  \Delta_t^{1,2}
  &=&2(3d_{t,P_{1}}+3d_{t,P_{4}}+d_{t,P_{7}})(3N_{t,P_{1}}+3N_{t,P_{4}}+N_{t,P_{7}})+N_{t,P_{1}}(3N_{t,P_{1}}+3N_{t,P_{4}}+N_{t,P_{7}})\nonumber \\
&\quad&+2(N_{t,P_{1}}+N_{t,P_{4}})(2N_{t,P_{1}}+N_{t,P_{4}})+N_{t,P_{1}}(N_{t,P_{4}}+N_{t,P_{7}})+(N_{t,P_{1}})^2.
\end{eqnarray}
Analogously, we find
\begin{eqnarray}\label{cross11}
  \Delta_t^{1,3}
  &=&2(3d_{t,P_{1}}+3d_{t,P_{4}}+d_{t,P_{7}})(3N_{t,P_{1}}+3N_{t,P_{4}}+N_{t,P_{7}})+2(N_{t,P_{1}})^2+2N_{t,P_{1}}(3N_{t,P_{1}}+3N_{t,P_{4}}+N_{t,P_{7}})\nonumber \\
&\quad&+N_{t,P_{4}}(3N_{t,P_{1}}+3N_{t,P_{4}}+N_{t,P_{7}})+(N_{t,P_{1}}+2N_{t,P_{4}}+N_{t,P_{7}})(2N_{t,P_{1}}+N_{t,P_{4}}),
\end{eqnarray}
\begin{eqnarray}\label{cross12}
  \Delta_t^{1,4}
  =2(3d_{t,P_{1}}+3d_{t,P_{4}}+d_{t,P_{7}})(3N_{t,P_{1}}+3N_{t,P_{4}}+N_{t,P_{7}})+(3N_{t,P_{1}}+3N_{t,P_{4}}+N_{t,P_{7}})^2+3(N_{t,P_{1}})^2,
\end{eqnarray}
and
\begin{equation}\label{cross13}
 \sum_{j \in \Omega_t^{4}}d_{Aj}
  =(3d_{t,P_{1}}+3d_{t,P_{4}}+d_{t,P_{7}})+(3N_{t,P_{1}}+3N_{t,P_{4}}+N_{t,P_{7}})+N_{t,P_{1}}.
\end{equation}

Substituting equations~(\ref{cross10}), (\ref{cross11}),
(\ref{cross12}), and (\ref{cross13}) into equation (\ref{cross02}),
we the final expression for cross distances $\Delta_t$,
\begin{eqnarray}\label{cross14}
\Delta_t =\frac{1}{3844}&\Big
[&5766+2728\times7^t+30290\times49^t-10974
\left(-1+\sqrt{2}\right)^t\nonumber\\&-&9300 \sqrt{2}
\left(-1+\sqrt{2}\right)^t +3114 \left(-1+\sqrt{2}\right)^{2
t}\nonumber\\&+&1017 \sqrt{2} \left(-1+\sqrt{2}\right)^{2 t}+186
\left(-1-\sqrt{2}\right)^t \left(-59+50
\sqrt{2}\right)\nonumber\\&-&9 \left(-1-\sqrt{2}\right)^{2 t}
\left(-346+113 \sqrt{2}\right)+16368 t\times 49^t\Big ].
\end{eqnarray}

\subsubsection{Exact result for average path length}

With the above-obtained results and recursion relations, we now
readily calculate the sum of the shortest path lengths between all
pairs of nodes. Inserting equation~(\ref{cross14}) into
equation~(\ref{total02}) and using the initial condition $D_{2}
=717$, equation~(\ref{total02}) is solved inductively,

\begin{eqnarray}\label{total04}
D_{t} =\frac{1}{161448}&\bigg
[&201810+215270\times7^t+11194\times7^{2 t}\nonumber\\&+&72072
\left(-1-\sqrt{2}\right)^t-57834 \sqrt{2}
\left(-1-\sqrt{2}\right)^t\nonumber\\&-&44037
\left(-1-\sqrt{2}\right)^{2t}-11340 \sqrt{2}
\left(-1-\sqrt{2}\right)^{2t}\nonumber\\&+&72072
\left(-1+\sqrt{2}\right)^t+57834 \sqrt{2}
\left(-1+\sqrt{2}\right)^t\nonumber\\&-&44037
\left(-1+\sqrt{2}\right)^{2 t}+11340 \sqrt{2}
\left(-1+\sqrt{2}\right)^{2 t}\nonumber\\&+&16368t\times 7^t
+16368t\times 7^{2 t} \bigg].
\end{eqnarray}

\begin{figure}
\begin{center}
\includegraphics[width=.25\linewidth,trim=80 30 80 30]{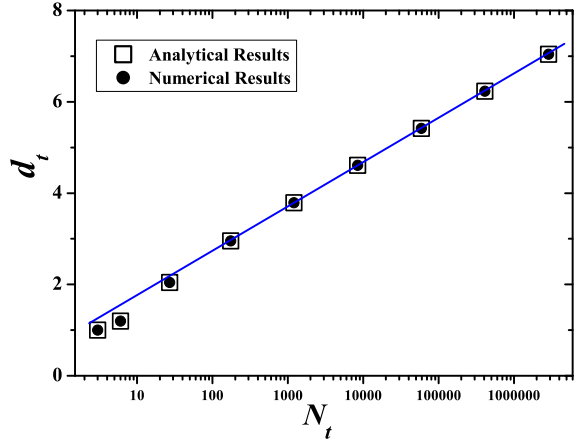}
\end{center}
\caption[kurzform]{ Average distance $d_{t}$ versus network order
$N_{t}$ on a semilogarithmic scale. The solid line serves as guide
to the eye.}\label{AveDis}
\end{figure}

Substituting equation~(\ref{total04}) into  equation~(\ref{apl01})
yields the exactly analytic expression for average path length
\begin{eqnarray}\label{apl02}
d_{t} =&\quad&\frac{1}{20181 (15+8\times7^t+7^{2t})}\bigg
[201810\nonumber\\&+&215270\times7^t+11194\times7^{2
t}\nonumber\\&+&72072 \left(-1-\sqrt{2}\right)^t-57834 \sqrt{2}
\left(-1-\sqrt{2}\right)^t\nonumber\\&-&44037
\left(-1-\sqrt{2}\right)^{2t}-11340 \sqrt{2}
\left(-1-\sqrt{2}\right)^{2t}\nonumber\\&+&72072
\left(-1+\sqrt{2}\right)^t+57834 \sqrt{2}
\left(-1+\sqrt{2}\right)^t\nonumber\\&-&44037
\left(-1+\sqrt{2}\right)^{2 t}+11340 \sqrt{2}
\left(-1+\sqrt{2}\right)^{2 t}\nonumber\\&+&16368t\times 7^t
+16368t\times 7^{2 t} \bigg].
\end{eqnarray}
In the large $t$ limit, $d_{t}\sim t$, while the network order $N_t
\sim 7^t$ which is obvious from equation~(\ref{Nt}). Thus, the
average path length grows logarithmically with increasing order of
the network. We have checked our analytic result provided by
equation~(\ref{apl02}) against numerical calculations for different
network order up to $t=8$ which corresponds to $N_{8}=1\,007\,772$.
In all the cases we obtain a complete agreement between our
theoretical formula and the results of numerical investigation, see
figure~\ref{AveDis}.

Recently, it has been suggested that for random uncorrelated
scale-free networks (SFNs) with degree exponent $\gamma <3$ and
network order $N$, their average distance $d(N)$ behaves as a double
logarithmic scaling with $N$: $d(N)\sim \ln\ln
N$~\cite{CoHa03,ChLu02}. However, for the deterministic network
considered here, in despite of the fact that its degree exponent
$\gamma=1+\frac{\ln7}{\ln3} <3$, its average path length scales as a
logarithmic scaling with network order, showing a obvious difference
from that of the stochastic scale-free counterparts. The logarithmic
scaling of $d_{t}$ with $N_t$ for our network as well as the
Apollonian networks~\cite{ZhChZhFaGuZo08} shows that previous
relation between APL and the network order obtained for uncorrelated
SFNs~\cite{CoHa03,ChLu02} is not valid for disassortative
SFNs~\cite{CoHa03,ChLu02}, at least for some spatial networks, e.g.,
the Apollonian networks and the network considered here. This leads
us to the conclusion that degree exponent itself does not suffice to
characterize the APL of SFNs.

\subsection{Degree correlations}

An interesting quantity related to degree correlations is the
average degree of the nearest neighbors for nodes with degree $k$,
denoted as $k_{\rm nn}(k)$, which is a function of node degree
$k$~\cite{PaVaVe01,VapaVe02}. When $k_{\rm nn}(k)$ increases with
$k$, it means that nodes have a tendency to connect to nodes with a
similar or larger degree. In this case the network is defined as
assortative \cite{Newman02}. In contrast, if $k_{\rm nn}(k)$ is
decreasing with $k$, which implies that nodes of large degree are
likely to have near neighbors with small degree, then the network is
said to be disassortative. If correlations are absent, $k_{\rm
nn}(k)=const$.

We can exactly calculate $k_{\rm nn}(k)$ for network $F_t$ using
equations~(\ref{deltak}) and (\ref{ki}) to work out how many links
are made at a particular step to nodes with a particular degree. By
construction, we have the following
expression~\cite{DoMa05,ZhRoZh07}
\begin{eqnarray}\label{Knn1}
k_{\rm nn}(k)&=&{1\over L_v(t_i) k(t_i,t)}\Bigg[
  \sum_{t'_i=0}^{t'_i=t_i-1} 2 L_v(t'_i) L_\Delta(t'_i,t_i-1)k(t'_i,t) \nonumber\\
 &\qquad&+\sum_{t'_i=t_i+1}^{t'_i=t} 2L_v(t_i) L_\Delta(t_i,t'_i-1)
 k(t'_i,t)\Bigg]+2
\end{eqnarray}
for $k=4\times 3^{t-t_i}$ ($t_i\geq1$), where $k(t_i,t)$ is the
degree of a node $i$ at time $t$ that was born at step $t_i$. Here
the first sum on the right-hand side accounts for the links made to
nodes with larger degree (i.e.\ $t'_i<t_i$) when the node was
generated at $t_i$. The second sum describes the links made to the
current smallest degree nodes at each step $t'_i>t_i$. The last term
2 accounts for the two links connected to two simultaneously
emerging nodes. After some algebraic manipulations, we can rewrite
equation~(\ref{Knn1}) in term of $k$ to obtain
\begin{eqnarray}\label{knn2}
k_{\rm nn}(k)=
\frac{(3^{2\,t+1}+3^{t-1})\left(\frac{4}{k}\right)^{2-\ln7/\ln3}}{2\times7^{t-1}}+\frac{8\ln(\frac{k}{4})}{3\ln3}-10.
\end{eqnarray}

For $k=3^{t}+1$ ($t_i=0$), we have
\begin{eqnarray}\label{Knn3}
k_{\rm nn}(k)&=&{1\over k(t_i,t)}\Bigg[
  \sum_{t'_i=t_i+1}^{t'_i=t} 2 L_\Delta(t_i,t'_i-1)
 k(t'_i,t)\Bigg]+2\nonumber\\
 &=&\frac{8t\times3^{t}}{3^{t+1}+3}+2.
\end{eqnarray}

Therefore, for large $t$ and $k$, $k_{\rm nn}(k)$ is approximately a
power law function of $k$ as $k_{\rm nn}(k)\sim k^{-\omega}$ with
$\omega=2-\frac{\ln7}{\ln3}\simeq 0.229$, which shows that the
network is disassortative. Note that $k_{\rm nn}(k)$ of the Internet
exhibits a similar power-law scaling with exponent $\omega=0.5$
\cite{PaVaVe01}.

\section{Conclusion}

In summary, motivated by the disk packing and Apollonian networks,
we have presented a model for spatial planar networks introducing
the influence of geography encoded in the disk packing. According to
the construction, we have studied analytically the main structural
features of the network. We have shown that the network has a power
law distribution with exponent $\gamma=1+\frac{\ln7}{\ln3}$, it has
a large clustering coefficient 0.603, its APL scales logarithmically
with the number of network nodes, and it is disassortative with the
average degree of the nearest neighbors for nodes having degree $k$
being roughly a power-law function of $k$ with exponent -0.229.

Note that although both the network considered here and the
Apollonian network~\cite{AnHeAnSi05,DoMa05} are translated from disk
packings, and both networks have qualitatively similar topologies,
their structural characteristics are quantitatively different. For
example, the exponent of degree distribution is
$1+\frac{\ln7}{\ln3}$ for our network, while for Apollonian network
it is $1+\frac{\ln3}{\ln2}$; the average clustering coefficients for
our network and Apollonian network are 0.603 and 0.828,
respectively. In addition, the average path
length~\cite{ZhChZhFaGuZo08} and the degree
correlations~\cite{DoMa05} for both networks are also of
quantitative difference. These disparities of the two networks show
that the ways of disk packing lead to different spatial constrains
of network nodes, which in turn have a significant impact on network
properties and thus dynamics running on networks, such as cascading
failing~\cite{HuYaYa06}, random walks~\cite{ZhGuXiQiZh09}, and so
on. Thus, we can conclude that for spatial networks the positions,
where nodes are geographically located, matter greatly and should be
incorporated when modeling such networks. Ignoring the geography
will lead to miss some important attributes and properties of the
systems.

\section*{Acknowledgment}

We would like to thank Yichao Zhang and Ming Yin for their help.
This research was supported by the National Basic Research Program
of China under grant No. 2007CB310806, the National Natural Science
Foundation of China under Grant Nos. 60704044, 60873040 and
60873070, Shanghai Leading Academic Discipline Project No. B114, and
the Program for New Century Excellent Talents in University of China
(NCET-06-0376).

\end{document}